\documentclass[5p,twocolumn]{elsarticle}
\usepackage[USenglish]{babel}
\usepackage[T1]{fontenc}
\usepackage[utf8]{inputenc}
\usepackage{amssymb} 
\usepackage{amsmath}
\usepackage{amsthm}
\usepackage{amsfonts}
\usepackage{siunitx}
\usepackage{booktabs}

\journal{Nuclear Instruments and Methods in Physics Research Section A}

\begin{document}


\title{Using short drive laser pulses to achieve net focusing forces in tailored dual grating dielectric structures}

\author[desy,uhh]{F.~Mayet}
\ead{frank.mayet@desy.de}

\author[desy]{R.~Assmann}
\author[desy]{U.~Dorda}
\author[desy,uhh]{W.~Kuropka}

\address[desy]{Deutsches Elektronen-Synchrotron DESY,
Notkestraße 85, 22607 Hamburg, Germany}
\address[uhh]{Universität Hamburg, Institut für Experimentalphysik,
Luruper Chaussee 149, 22761 Hamburg, Germany}

\begin{abstract}
Laser-driven grating type DLA (Dielectric Laser Accelerator) structures have been shown to produce accelerating gradients on the order of GeV/m. In simple $\beta$-matched grating structures due to the nature of the laser induced steady-state in-channel fields the per period forces on the particles are mostly in longitudinal direction. Even though strong transverse magnetic and electric fields are present, the net focusing effect over one period at maximum energy gain is negligible in the case of relativistic electrons. Stable acceleration of realistic electron beams in a DLA channel however requires the presence of significant net transverse forces. In this work we simulate and study the effect of using the transient temporal shape of short Gaussian drive laser pulses in order to achieve suitable field configurations for potentially stable acceleration of relativistic electrons in the horizontal plane. In order to achieve this, both the laser pulse and the grating geometry are optimized. Simulations conducted with the Particle-In-Cell code VSim 7.2 are shown for both the transient and steady state/long pulse case. Finally we investigate how the drive laser phase dependence of the focusing forces could affect a potential DLA-based focusing lattice. 
\end{abstract}

\begin{keyword}
	dielectric laser accelerator \sep simulation \sep optimization \sep focusing \sep ACHIP
\end{keyword}

\maketitle



\section{Introduction}
\begin{figure}[!htb]
	\centering
	\includegraphics[width=0.45\textwidth]{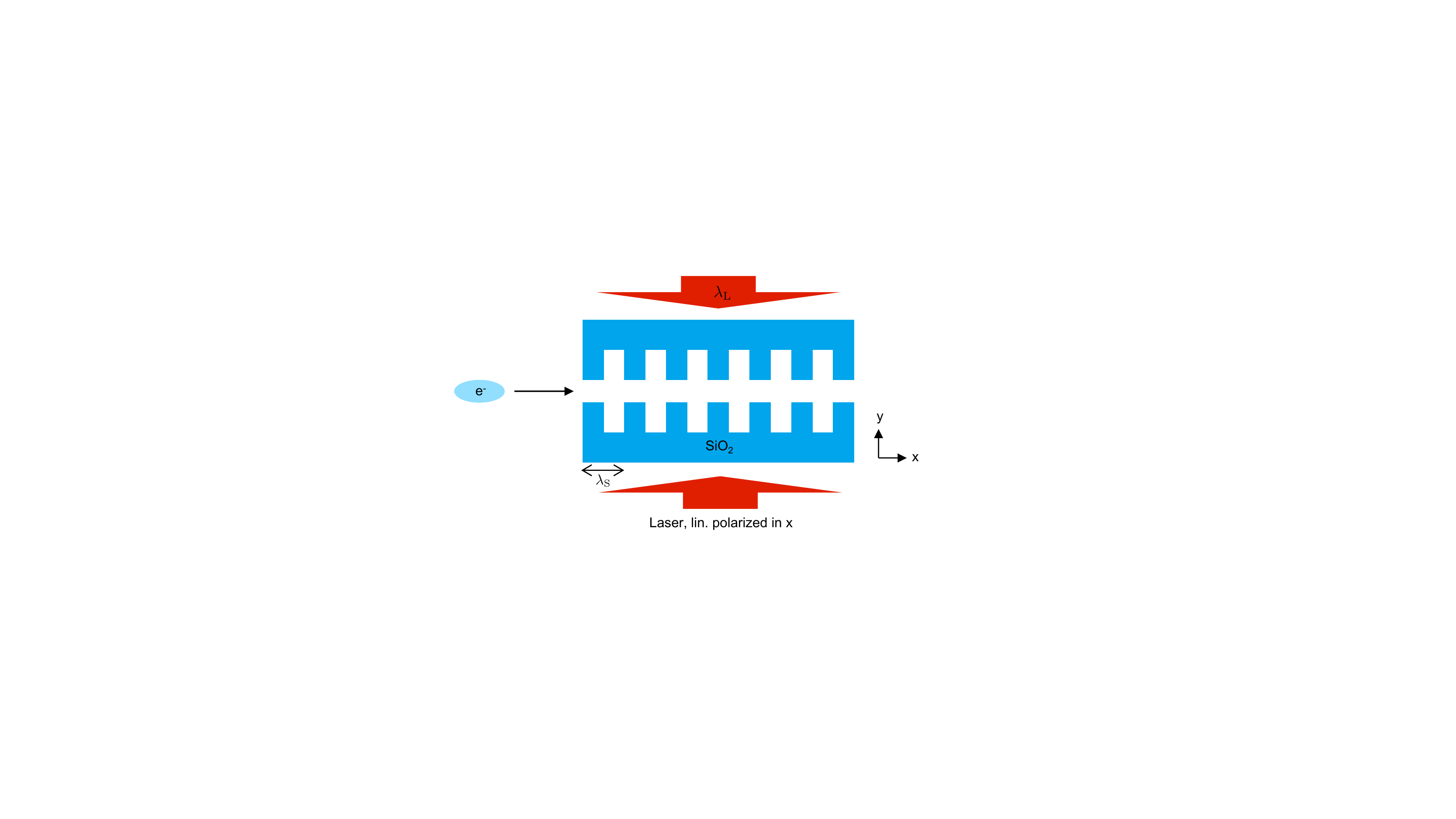}
	\caption{Schematic of a dual grating DLA illuminated from both sides with a linearly polarized laser field with wavelength $\lambda_\text{L}$. $\lambda_\text{S}$ is the period length, which is connected to the laser wavelength by the \emph{synchronicity condition} $\lambda _\text{S} = \beta _\text{m} \lambda _\text{L}$, where $\beta _\text{m}$ is the matched normalized particle velocity. The electrons travel along $x$ with normalized velocity $\beta _\text{p}$.}
	\label{Fig1}
\end{figure}

The concept of dielectric laser accelerators (DLA) has gained increasing attention in accelerator research, because of the high achievable acceleration gradients ($\sim$GeV/m) \cite{England:2014bf}. This is due to the high damage threshold of dielectrics at optical frequencies. One of the structure types being studied is the symetrically driven dual grating. A schematic of this type of structure is shown in Fig.~\ref{Fig1}. 
If the structure periodicity $\lambda _\text{S}$ and the drive laser wavelength $\lambda _\text{L}$ are adjusted according to the normalized velocity $\beta _\text{p}$ of the injected electrons, phase-synchronous acceleration can be achieved. In this case the structure is then called $\beta$-matched. In simple $\beta$-matched grating structures due to the nature of the laser induced in-channel fields the steady-state per period longitudinal force on the particle is $\pi / 2$ out of phase with the transverse force (see Sec. \ref{sec:AnalyticalDescription}). Therefore at maximum energy gain phase no transverse forces are exerted on the particle. Stable acceleration and transport of realistic electron beams in a DLA channel however requires the presence of significant net transverse forces. In this work we simulate and study how a) the magnitude of the transverse force can be enhanced and b) how the phase relation between the longitudinal and transverse force can be altered by tailoring the structure geometry and hence leaving the ideal grating picture as shown in Fig.~\ref{Fig1}. In the final section (Sec. \ref{sec:DLATransportLine}) we investigate how the drive laser phase dependence of the resulting focusing forces could affect a potential future DLA-based focusing lattice.
\section{Analytical Description of the Fields} \label{sec:AnalyticalDescription}
For the following calculation an in x-direction linearly polarized plane wave, which is traveling along the y-direction with a wavelength of $\lambda _0$ is assumed (cf. Fig.~\ref{Fig1}). It is incident on a \emph{single} grating perpendicular to the grating structure, which implies a structure being periodic in x-direction. The problem is assumed to be pseudo-2-dimensional in the sense that in z-direction the structure is infinite.

Using \emph{Maxwell's Equations} we see that our plane wave has a non-vanishing magnetic field only in z-direction. The magnetic field after passage of the grating can be described as a composition of an infinite number of \emph{spatial harmonics}, or diffraction modes (cf. \emph{Floquet Theorem} $\rightarrow$ Fourier series) \citep{Breuer-0953-4075-47-23-234004}:
\[
	B_z(x,y,t) = \sum _{n=-\infty}^\infty B_{z,0}^{(n)} \cdot e^{i(nk_x x + k_y y -\omega t + \phi _0)},
\]
where $k_x = 2\pi / \lambda _x$ is the wave vector component w.r.t the grating period and $B_{z,0}^{(n)}$ is the $n$-th order complex Fourier weight. The term $\phi _0$ is an arbitrary phase offset. Here it describes the particle to laser phase relation. Inserting $B_z$ into the wave equation for vacuum it can be seen that $k_y ^2 = k_0 ^2 - n^2k_x^2$. Hence
\[
  B_z(x,y,t) = \sum _{n=1}^\infty B_{z,0}^{(n)} \cdot e^{i\left(y\sqrt{k_0^2 - n^2k_x^2} + nk_x x - \omega t + \phi _0\right)}.
\]
Using $\nabla \times \vec B = -i \cdot \frac{k_0}{c} \vec E$ the x and y components of the electric field can be calculated from $B_z$. In order to efficiently accelerate a moving particle it has to be phase synchronous with the parallel component of the electromagnetic field. From this it follows that
\begin{equation}
  \frac{\omega}{nk_x} = \beta _\text{m} c \Leftrightarrow k_x = \frac{k_0}{n \beta _\text{m}} \Leftrightarrow \beta _\text{m} = \frac{\lambda _x}{n \lambda _0},
\label{eq:kx}
\end{equation}
where $k_0$ is the laser wave number and $\beta _\text{m} c$ is the \emph{matched} particle velocity. Eq.~(\ref{eq:kx}) is the \emph{synchronicity condition} for grating accelerators for the $n$-th harmonic. In the following calculations we use $n=1$ as the synchronous order and hence $k_x = k_0/\beta _\text{m}$. Inserting $k_x$ into the expressions for the field components and taking only the real part yields
\begin{equation}
  \begin{split}
  \Re(\mathbf{B})(x,y,t) &= \sum _{n=-\infty}^\infty |B_{z,0}^{(n)}| \cdot e^{-\delta _\text{m}^{(n)} y}\\
  &\cdot
  \begin{pmatrix}
    0\\
    0\\
    \cos{\left(n\frac{k_0}{\beta _\text{m}} x - \omega t + \tilde \phi _0^{(n)}\right)}
  \end{pmatrix},\\
  \Re(\mathbf{E})(x,y,t) &= \sum_{n=-\infty}^{\infty} |E_{x,0}^{(n)}| \cdot e^{-\delta _\text{m} ^{(n)} y}\\
  &\cdot
  \begin{pmatrix}
    \sin \left(n\frac{k_0}{\beta _\text{m}} x - \omega t + \tilde \phi _0^{(n)}\right)\\
    \sqrt{\frac{n^2}{n^2 - \beta _\text{m}^2}} \cdot \cos \left(n\frac{k_0}{\beta _\text{m}} x - \omega t + \tilde \phi _0^{(n)}\right)\\
    0
  \end{pmatrix},
\end{split}
\label{eq:realfields}
\end{equation}
where $\delta _\text{m}^{(n)} = k_0 \sqrt{\frac{n^2}{\beta _\text{m}^2} - 1}$, $\tilde \phi _0^{(n)} = \phi _0 + \phi^{(n)}$ and $|E_{x,0}^{(n)}| = c \sqrt{\frac{n^2}{\beta _\text{m}^2} - 1} \cdot |B_{z,0}^{(n)}|$. $|B_{z,0}^{(n)}|$ and $\phi^{(n)}$ are the amplitudes and phases of the complex weights of the spatial harmonics respectively. It can be seen that the field falls off exponentially in y. The particles are accelerated in evanescent field components, which is required by the \emph{Lawson-Woodward Theorem}. From Eq.~(\ref{eq:realfields}) it can already be seen that in the case of an ideal grating DLA the longitudinal and transverse fields are out of phase by $\pi/2$. In order to describe a dual grating DLA two grating fields need to be superimposed, while taking both the channel width $L_\text{gap}$ and the opposite travel direction of the second drive laser into account.
\section{Per-Period Force on the Particle}
In the previous section the \emph{steady state} in-channel fields of a single and dual grating DLA were described. Using this field description it is now possible to estimate the per-period effect on a particle traversing the channel along the grating grooves with longitudinal velocity $\beta _\text{p} c$. For now we consider a zero-emittance single particle. It is also assumed for simplicity that $\beta _\text{p} = \text{constant}$ during the traversal of a single grating period.

First it is necessary to map the time-dependence of the field to the particle position along the channel. Since $x = \beta _\text{p} c t$ and $\omega = ck_0$, we can substitute $\omega t \Rightarrow k_0 x/\beta _\text{p}$. It is now helpful to introduce the particle to grating spatial phase $\Psi_\text{m}^{(n)}(x)$, which is defined as
\begin{equation}
	\Psi_\text{m}^{(n)}(x) = k_0 \left( \frac{n}{\beta _\text{m}} - \frac{1}{\beta _\text{p}} \right)x.
	\label{eq:spatialphase}
\end{equation}
In the $\beta$-matched case it reduces to
\[
	\Psi^{(n)}(x) = \frac{k_0}{\beta _\text{m}} ( n - 1 )x.
\]
Note that Eq.~(\ref{eq:spatialphase}) is zero for $n=1$ in the $\beta$-matched case, which eliminates the spatial dependence of the force on the electron during traversal of the period, which is just a re-statement of the phase-synchronicity of the first order.
In order to simplify the equations it is furthermore helpful to define $\Delta _{\text{m},\pm}^{(n)}(y)$, the transverse decay factor, which is defined for the dual grating case as
\begin{equation}
 	\Delta _{\text{m},\pm}^{(n)}(y) = e^{-\delta_\text{m}^{(n)}y} \pm e^{-\delta_\text{m}^{(n)}(L_\text{gap} - y)}
 	\label{eq:decayfunction}
\end{equation} 
and for the single grating case as
\[
 	\Delta _\text{m}^{(n)}(y) = e^{-\delta_\text{m}^{(n)}y}.
\]
The force on a moving particle caused by the presence of an electromagnetic field is given by the \emph{Lorentz Force} $\mathbf{F}_\text{L} = q (\mathbf{E} + \mathbf{v} \times \mathbf{B})$, where $q$ is the particle charge and $\mathbf{v}$ its velocity. Therefore the non-zero force components are in our case given by $F_x = qE_x$ and $F_y = q(E_y - v_x B_z)$, where $v_x = \beta _\text{p} c$. Inserting the expressions for $E_x$ and $E_y$, as well as the newly introduced definitions $\Psi_\text{m}^{(n)}(x)$, $\Delta _{\text{m},\pm}^{(n)}(y)$ and $\tilde \phi _0^{(n)}$ yields for $n \neq 0$
\begin{equation}
	\label{eq:forces}
	\begin{split}
		F_x^{(n)}(x,y) &= q c \sqrt{\frac{n^2}{\beta _\text{m}^2} - 1} \cdot B_{z_0}^{(n)} \cdot \sin (\Psi_\text{m}^{(n)}x + \tilde \phi _0^{(n)})\\ &\cdot \Delta _{\text{m},+}^{(n)}(y),\\
		F_y^{(n)}(x,y) &= q c \left[\frac{n}{\beta _\text{m}} - \beta _\text{p} \right] \cdot B_{z_0}^{(n)} \cdot \cos (\Psi_\text{m}^{(n)}x + \tilde \phi _0^{(n)})\\ &\cdot \Delta _{\text{m},-}^{(n)}(y).
	\end{split}
\end{equation}
The average per-period (or net-) force can now be calculated as
\[
	\langle F_{x,y} \rangle^{(n)}(y) = \frac{1}{\lambda _x} \int_0^{\lambda _x} F_{x,y}^{(n)}(x,y) dx
\]
where it is assumed that the transverse motion during one period is negligible and hence $ds \rightarrow dx$. It can be shown that in the $\beta$-matched case $\langle F_{x,y} \rangle^{(n)}(y) = 0$, $\forall n \neq 1$. This is not generally true anymore if $\beta _\text{m} \neq \beta _\text{p}$.
\subsection{Significant Per-Period Transverse Force}
As described above, in the case of an ideal dual grating DLA Eq.~(\ref{eq:forces}) can be used to estimate the per-period force on a particle traveling along the DLA channel. Taking a closer look on $F_y^{(n)}(x,y)$ it is possible to identify different ways to enhance the transverse force. Considering the term
\[
	\left[ \frac{n}{\beta _\text{m}} - \beta _\text{p} \right]
\]
two options come to mind:
\begin{itemize}
	\item Significant mismatch between $\beta _\text{m}$ and $\beta _\text{p}$
	\item \emph{Virtual} mismatch between $\beta _\text{m}$ and $\beta _\text{p}$ by injecting the electrons with an angle in the $x-z$ plane \citep{Plettner:2008bl}
\end{itemize}
At the same time if the grating is mismatched - as has been already stated above - higher orders do not cancel out anymore. It has to be noted however that the enhancement of the per-period transverse force via mismatch is not trivial since the spatial phase also changes with the level of mismatch (see Eq.~\ref{eq:spatialphase}). In addition to that, due to the increasingly fast decay of the higher orders (see Eq.~\ref{eq:decayfunction}) the accelerating field is much less homogeneous across the channel, which results in increased correlated energy spread.

Then there is a third option, which is based on letting the $B_{z,0}^{(n)}$ depend on $t$, or subsequently $x$. This means $B_{z,0}^{(n)} \rightarrow B_{z,0}^{(n)}(x)$, or in other words: The driving field envelope is not constant, which needs to be taken into account in the integration of the average per-period force and can lead to an enhancement as shown below.
\subsection{Using Time-dependent Fourier Weights}
The following calculations are based on four assumptions:
\begin{itemize}
	\item The slope of the $B_{z,0}^{(n)}(x)$ is linear over one period
	\item There is only negligible transverse motion of the electron over one period
	\item The electron to laser phase is constant over one period
	\item The structure is and stays $\beta$-matched
\end{itemize}
It is reasonable to assume that the slope of the field during one single grating period is almost linear and hence $B_{z,0}^{(n)}(x) = B_{z,0,0}^{(n)} + \text{d}_x B_{z,0}^{(n)} \cdot x$, where $\text{d}_x \rightarrow \text{d}/\text{d}x$. With this we get for $n \neq 0$
\begin{equation}
	\begin{split}
		\langle F_{y} \rangle^{(n)}(y) &\propto\\ 
		&\int_0^{\lambda _x}  (B_{z,0,0}^{(n)} + \text{d}_x B_{z,0}^{(n)} \cdot x) \\
		&\cdot \cos (\Psi^{(n)}x + \tilde \phi _0^{(n)}) dx.
	\end{split}
\end{equation}
In case of the synchronous mode $n=1$ this integral yields:
\begin{equation}
	\langle F_{y} \rangle^{(1)}(y) \propto \left( B_{z,0,0}^{(1)} + \frac{\text{d}_x B_{z,0}^{(1)}}{2} \cdot \lambda _x \right) \cdot \cos(\tilde \phi _0^{(1)}).
	\label{eq:transverseforcetransient}
\end{equation}
This means that the first order transverse force depends on both the amplitude and its slope. If the slope is sufficiently steep (since it is multiplied with the drive laser wavelength), the transverse force can be enhanced by this term. It is also interesting to note that if the slope term dominates the sign of the slope will cause a $\pi$ phase flip.
\begin{figure}[!htb]
	\centering
	\includegraphics[width=0.45\textwidth]{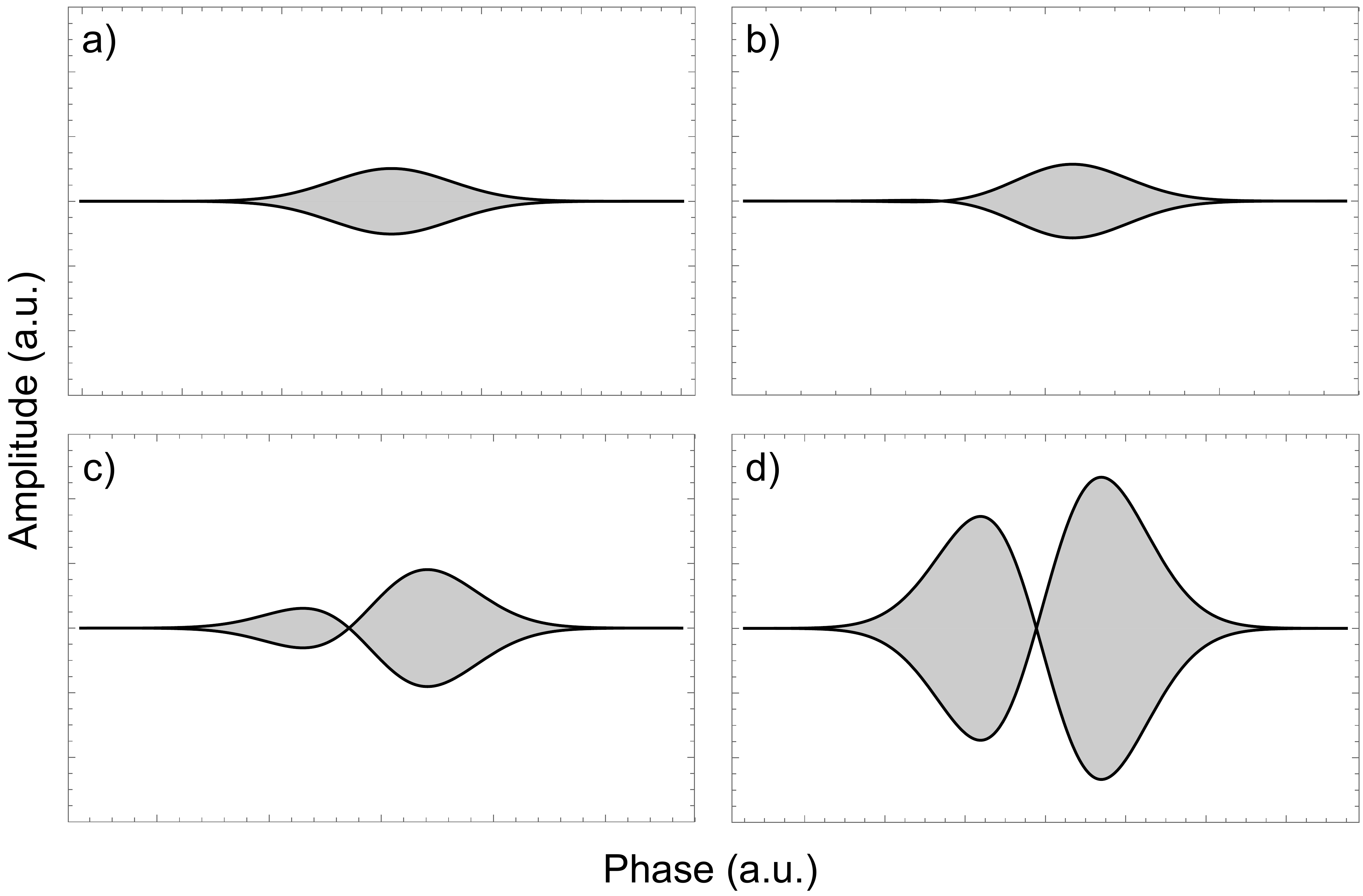}
	\caption{Illustration of the resulting transverse per-period force envelope according to Eq.~\ref{eq:transverseforcetransient} using the time dependence as described by Eq.~\ref{eq:gaussian}. The pulse length is reduced from a) to d). The amplitude scale is kept constant and in a.u., while the phase scale is always given by the interval $[-5 \sigma, 5\sigma]$.}
	\label{Fig2abcd}
\end{figure}
In Fig.~\ref{Fig2abcd} the envelope of Eq.~\ref{eq:transverseforcetransient} is shown for different time dependent $B_{z,0}^{(1)}(t)$. The time dependence was chosen to be a Gaussian as defined by
\begin{equation}
	g(x) = \frac{1}{\sigma \sqrt{2\pi}} e^{-\frac{1}{2}\left( \frac{x-\mu}{\sigma} \right)^2},
	\label{eq:gaussian}
\end{equation}
where $\mu$ is the center and $\sigma$ is the rms width as usual. It can be seen that for long pulses the slope does not play a role yet and the resulting pulse shape is just the initial Gaussian. If however the pulse gets short enough and the slope term large enough, the pulse shape gets modified substantially as two peaks emerge. Compared to the initial pulse shape the peak amplitude is enhanced.
\section{Simulations}
\begin{figure}[!htb]
	\centering
	\includegraphics[width=0.45\textwidth]{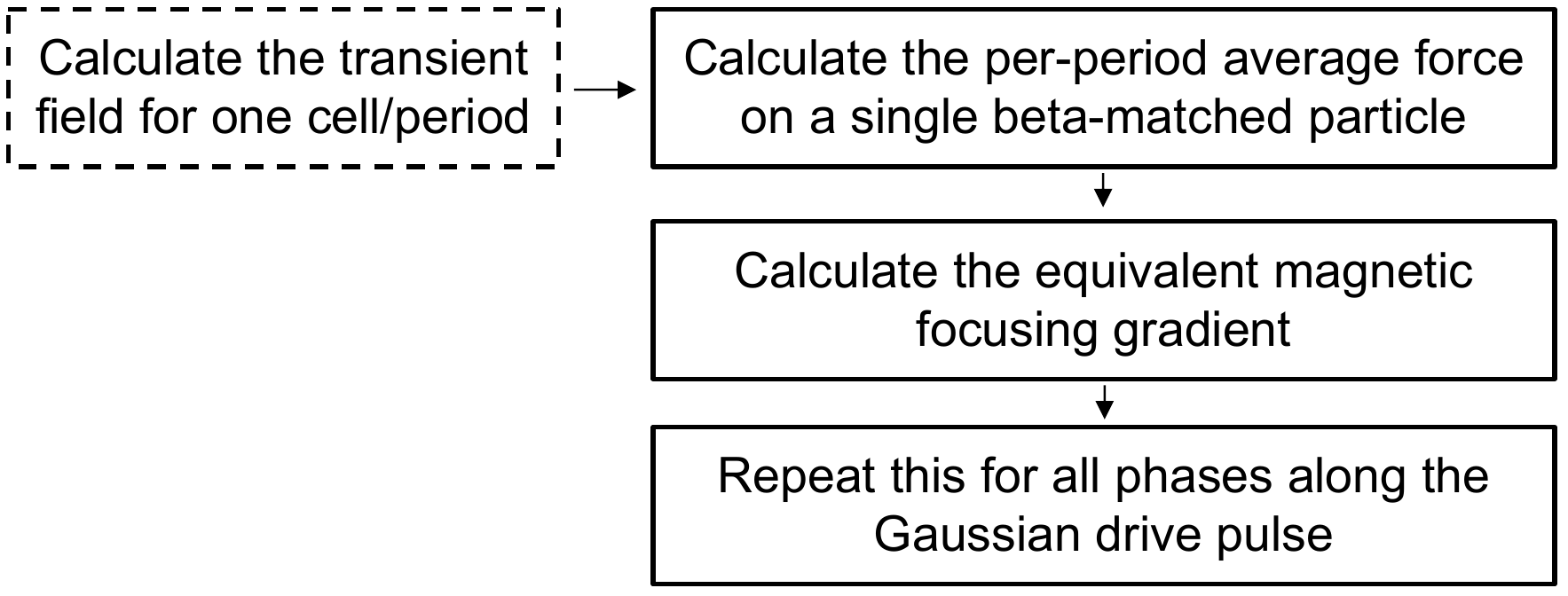}
	\caption{Simulation scheme for the numerical field simulations using VSim 7.2.}
	\label{FigScheme}
\end{figure}

In order to verify the force enhancement effect shown in the previous section numerical FDTD (Finite-Difference-Time-Domain) simulations were carried out using VSim 7.2 \citep{VSim}. Simulations were performed for both the steady state case and a short pulse case using Eq.~\ref{eq:gaussian} as the envelope, where the rms drive laser pulse length was chosen to be 50\,fs and the peak amplitude 0.5\,GV/m. The two drive lasers were assumed to be phase-locked and incident on a $\beta-$matched dual grating. $\beta _\text{m}$ was chosen to be close to 1 ($\rightarrow 50$\,MeV). As the figure of merit the equivalent magnetic focusing gradient as described in \citep{Wootton:2017kn} was used. It is defined by
\begin{equation}
	G = \frac{\partial _\perp \langle F_\perp \rangle _\lambda}{q_e c},
	\label{eq:G}
\end{equation}
where $\partial _\perp \langle F_\perp \rangle _\lambda$ is the derivative of the per-period average transverse force w.r.t. the transverse coordinate, $q_e$ the electron charge and $c$ the speed of light. The unit of $G$ is [T/m] as expected. The simulation was carried out according to the scheme shown in Fig.~\ref{FigScheme}. Fig.~\ref{Fig3} shows the results of the simulation and subsequent data processing. The solid line shows $G$ and the scatter plot the accelerating part of the longitudinal force, where the size of the data point symbolizes its magnitude. The decelerating part is omitted for readability. In comparison to the steady state both the two emerging peaks (as already seen for short pulses in Fig.~\ref{Fig2abcd}), as well as the enhancement of $G$ by a factor of $\sim 2.5$ at the maximum can be seen. The focusing gradient indeed seems to depend on the slope of the spatial harmonic weights as predicted by the simple calculations shown in the previous sections. It can also be seen that the phase relation between acceleration and focusing $\Phi (t)$ changes along the pulse.
\subsection{Acceleration to Focusing Phase}
In the previous section the possible enhancement of $G$ using short drive pulses was shown. Even though $\Phi (t) \neq \text{const.}$ along the pulse it is never flat 0. If this was the case, the beam would be focused and accelerated at the same time on-crest.
\setcounter{figure}{4}
\begin{figure}[!ht]
	\centering
	\includegraphics[width=0.45\textwidth]{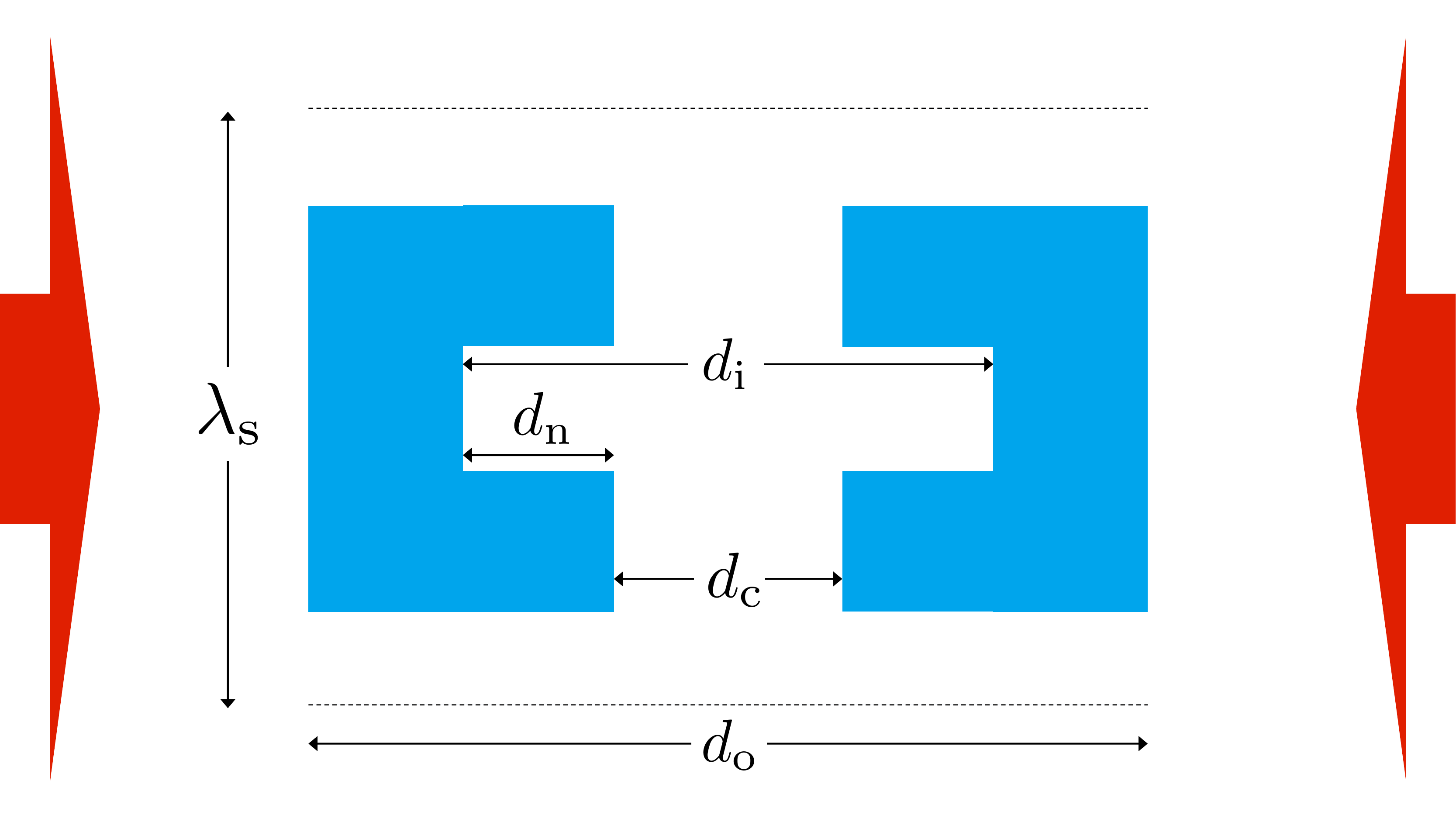}
	\caption{The new structure design. Only one cell is shown.}
	\label{FigStruct}
\end{figure}

\begin{figure}[!htb]
	\centering
	\includegraphics[width=0.45\textwidth]{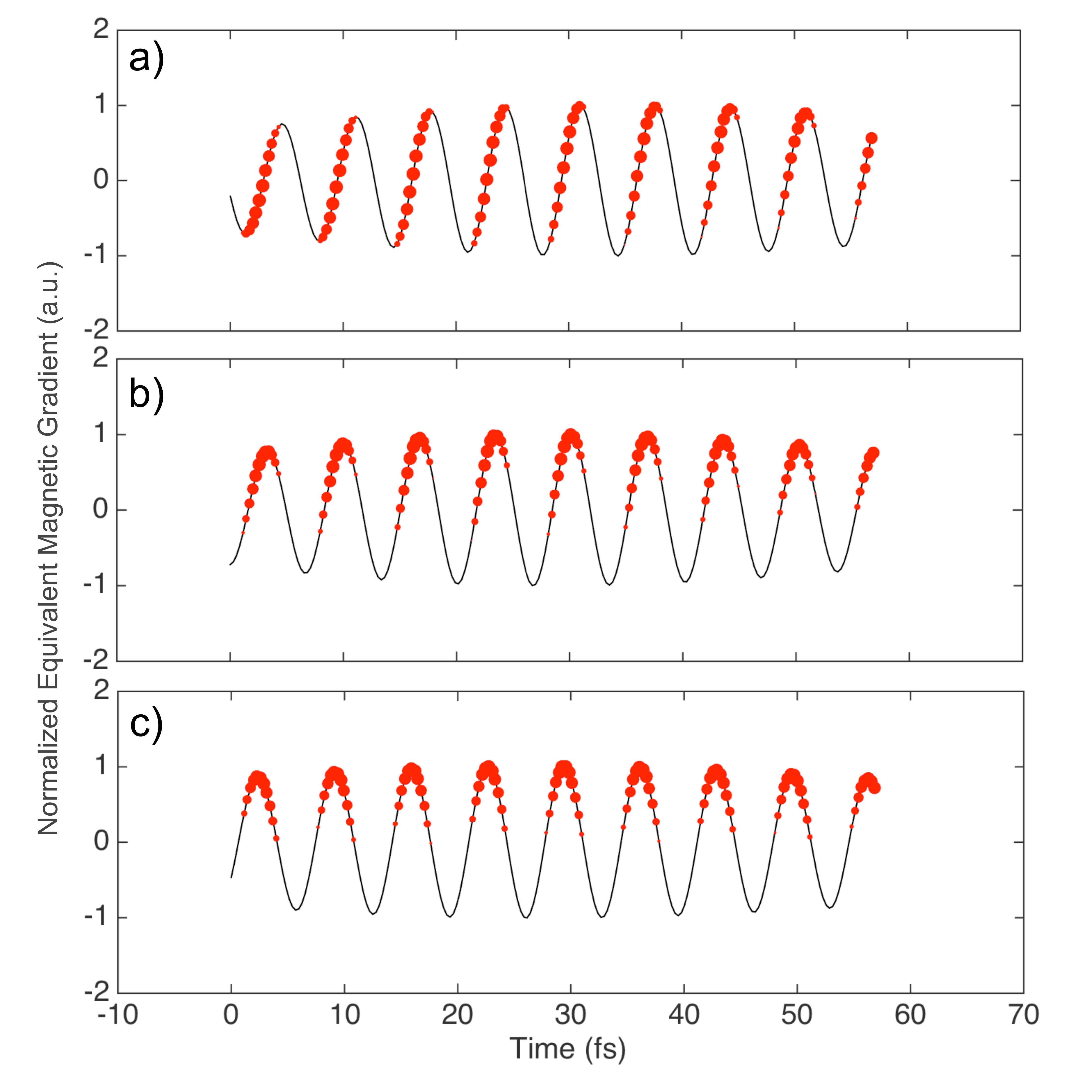}
	\caption{Normalized equivalent magnetic focusing gradient vs. time. The scatter plot shows the accelerating part of the longitudinal force, where the size of the data points symbolizes its magnitude. From a) to c) both $d_\text{n}$ and $d_\text{c}$ are decreased slighty.}
	\label{Fig5}
\end{figure}

In order to achieve this, we investigated a modified grating structure as shown in Fig.~\ref{FigStruct}. Since this structure -- due to its shape -- adds a certain amount of resonance to the system the analytical description as shown above does not apply completely anymore. Hence we also expect $\Phi (t)$ to evolve differently along the pulse. Fig.~\ref{Fig5} shows $G$ in a section of the pulse around its peak for three different values of the geometry shape $(d_\text{n}, d_\text{c})$.
It can be seen that $\Phi (t)$ can indeed be altered to a degree that $\Phi (t) \approx 0$ over several periods of the pulse. 
\setcounter{figure}{3}
\begin{figure*}[!h]
	\centering
	\includegraphics[width=1.0\textwidth]{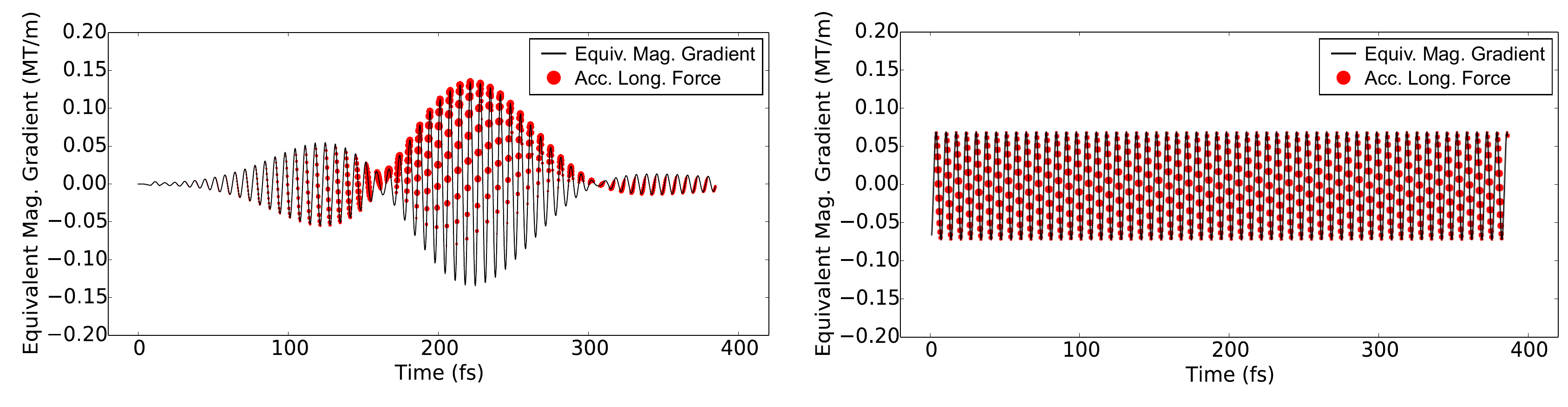}
	\caption{Equivalent magnetic focusing gradient $G$ vs. time, calculated from VSim 7.2 field data (solid line). The scatter plot shows the accelerating part of the longitudinal force, visualizing the change of the phase between acceleration and focusing along the pulse. The size of the data points symbolizes the magnitude of the force. \textbf{Left}: \SI{50}{\femto\second} rms Gaussian drive pulse. \textbf{Right}: Steady state case.}
	\label{Fig3}
\end{figure*}
\setcounter{figure}{6}
Interestingly in this configuration $d_\text{o} = 2 \lambda_\text{L}, d_\text{i} = \lambda_\text{L}$. It has to be noted that the peak amplitude of $G$ is also altered.

\begin{figure}[!htb]
	\centering
	\includegraphics[width=0.45\textwidth]{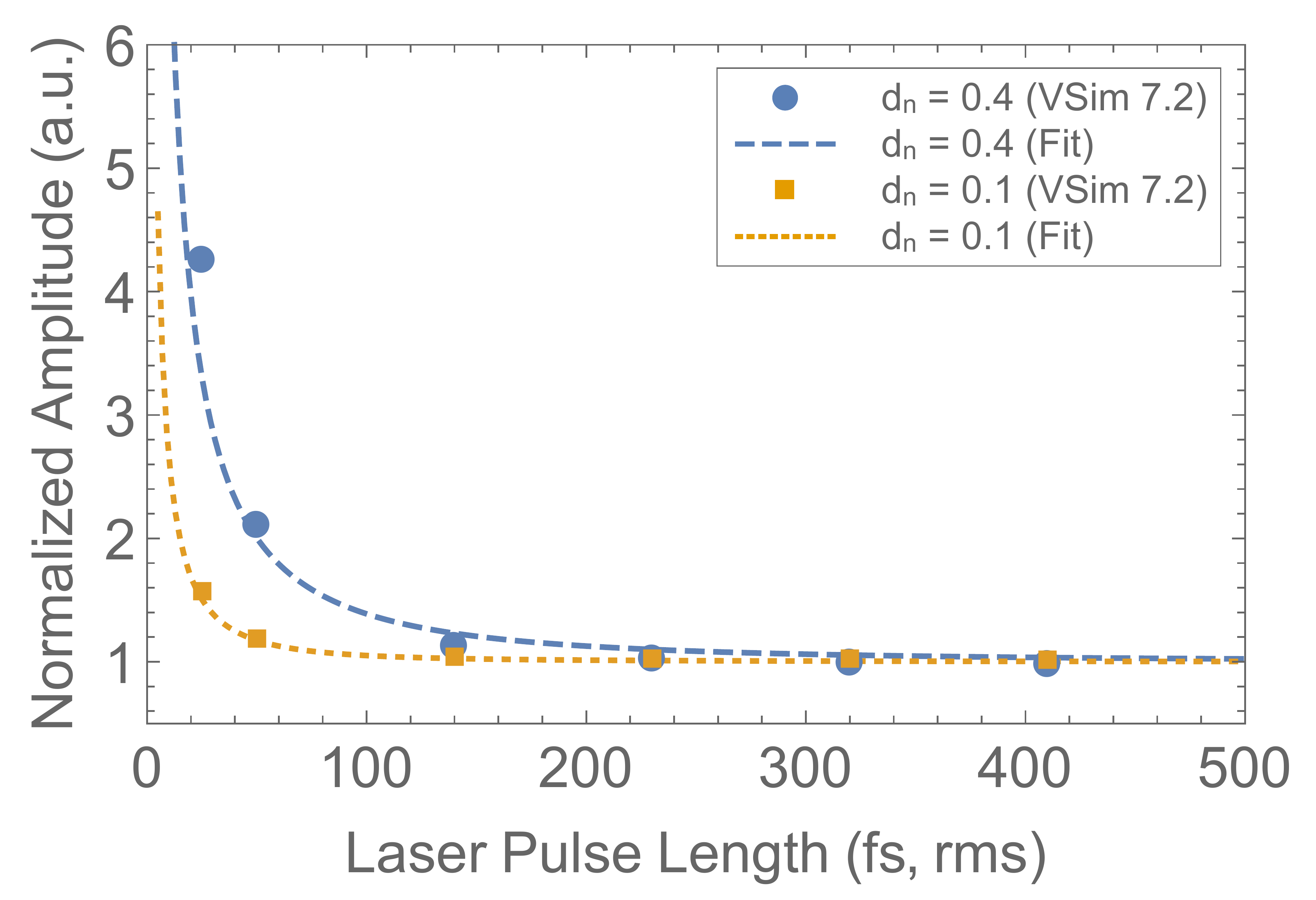}
	\caption{Normalized maximum achieved $G$ for different rms drive laser pulse lengths simulated with VSim 7.2, as well as a fit to Eq.~\ref{eq:G}. Two exemplary geometries defined by $d_\text{n}$ are shown.}
	\label{FigLaserPulseLengthScan}
\end{figure}

Fig.~\ref{FigLaserPulseLengthScan} shows the maximum achieved $G$, normalized to the steady state result, for different rms drive laser pulse lengths, as well as a fit to Eq.~\ref{eq:G}, using Eq.~\ref{eq:transverseforcetransient} for $F_\perp$ and a free scaling factor for $\sigma$ in Eq.~\ref{eq:gaussian}. It can be seen that with decreasing pulse length the maximum $G$ is enhanced as discussed in the last section. It can also be seen that the onset of the enhancement effect is influenced by the structure geometry as well.
\section{DLA Transport Line} \label{sec:DLATransportLine}
If in the future a fully integrated DLA-based accelerator should be able to reach high energies, a need for beam transportation along the miniaturized beamline is implied. In conventional accelerators transport lattices often comprise multiple so called FODO cells, which combine a focusing and a defocusing quadrupole separated by drift sections. This is due to the fact that quadrupoles can only focus in one plane. In the respective other plane the beam is defocused. By alternating the focusing plane the beam can be kept stable. 

Since a simple grating type DLA only acts on one transverse plane, it will also be necessary to build a kind of FODO lattice out of DLAs, where every second device is rotated by 90\,deg. A similar setup has already been studied by Kuropka et al. \cite{Kuropka:IPAC2017-WEPVA005}. Here we focus on the fact that due to the short period length $\lambda _\text{p}$ the equivalent quadrupole focusing strength $k$ strongly depends on the longitudinal position of a particle within the channel (or laser-to-electron phase $\phi _0$). It is worth noting that over a phase range of $\pi/2$ the equivalent $k$ can vary many orders of magnitude, even going down to 0. For a finite length beam this phase dependence has a strong effect on the betatron phase advance $\Psi_{x,y}$ of the individual slices along the beam.
\subsection{Phase dependent Transport Matrix}
The goal of this section is to write down a transfer matrix for the particle phase space coordinate $(x,x^\prime)$ in analogy to a quadrupole transfer matrix, with the inclusion of a phase dependent $k$. The well known quadrupole transfer matrix is given by
\begin{equation}
	\begin{split}
	\mathbf{M}_\text{QF} &= 
	\begin{pmatrix}
		\cos(\Omega) & \frac{1}{\sqrt{k}} \sin(\Omega)\\
		-\sqrt{k} \sin(\Omega) & \cos(\Omega)
	\end{pmatrix},\\
	\mathbf{M}_\text{QD} &= 
	\begin{pmatrix}
		\cosh(\Omega) & \frac{1}{\sqrt{|k|}} \sinh(\Omega)\\
		\sqrt{|k|} \sinh(\Omega) & \cosh(\Omega)
	\end{pmatrix},
	\end{split}
\end{equation}
where $\Omega = \sqrt{|k|}s$ and $s$ is the quadrupole length. If $k > 0$ the quadrupole is focusing ($\mathbf{M}_\text{QF}$), if $k < 0$ it is defocusing ($\mathbf{M}_\text{QD}$). Otherwise it is just a drift section. The $k$ value is usually calculated via
\[
	k = \frac{e}{p} \cdot \frac{dB_z}{dy},
\]
where $e$ is the electron charge and $p = \sqrt{E^2/c^2 - m_0^2c^2}$ the total momentum of the particle. In order to incorporate the phase dependence inside the DLA, we re-define it in accordance with \citep{Wiedemann:2007ws} as
\begin{equation}
	k(\phi_0) = 2.998 \cdot \frac{G}{\beta _\text{p}E[\text{GeV}]} \cdot \cos (\phi _0),
\end{equation}
where $G$ is the equivalent magnetic gradient as defined by Eq.~\ref{eq:G}. It is now possible to construct the FODO lattice by matrix multiplication of focusing, defocusing and drift matrices respectively.
\subsection{Simulation}
\begin{figure}[!htb]
	\centering
	\includegraphics[width=0.45\textwidth]{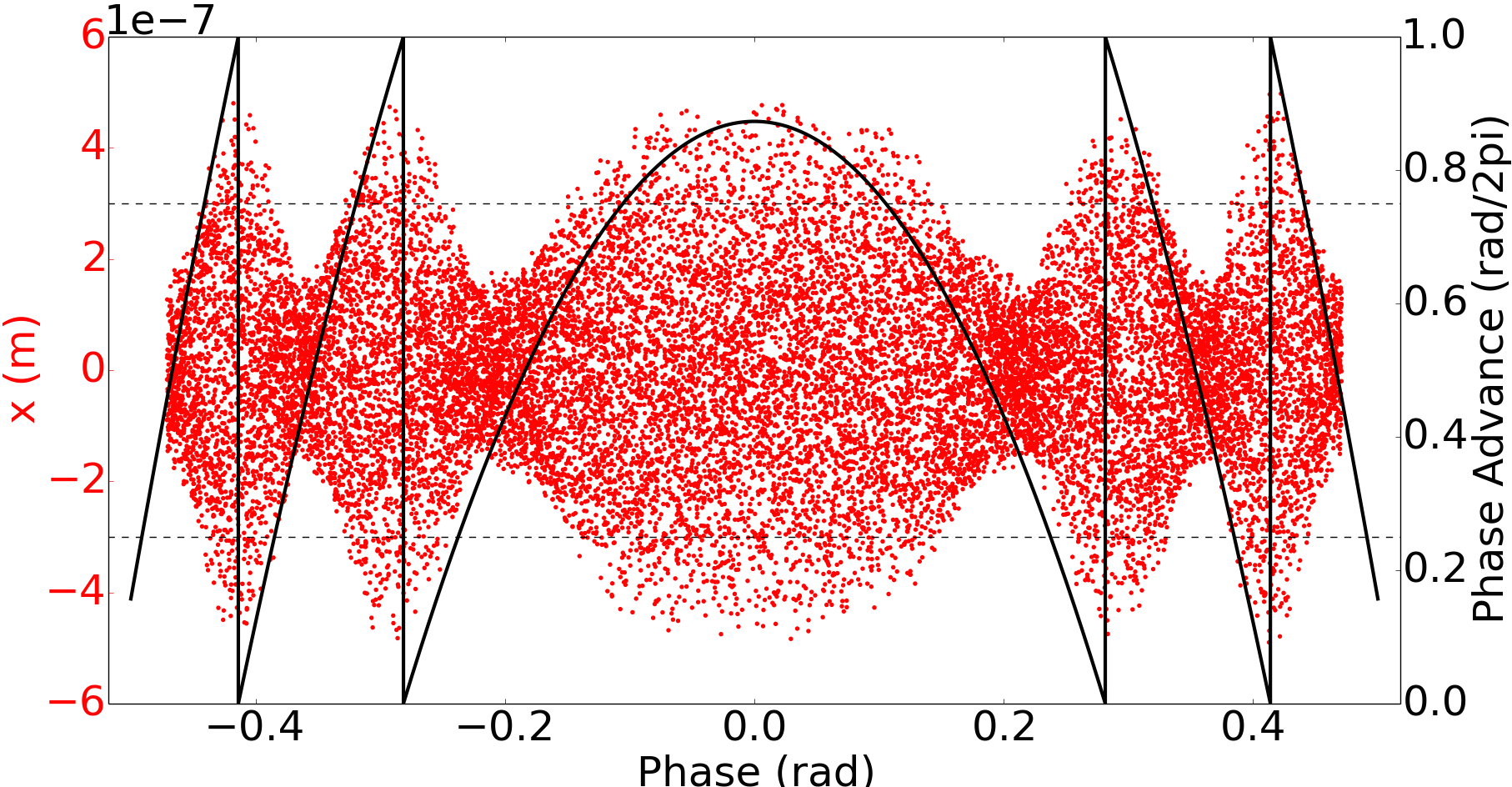}
	\caption{The beam after 1000 FODO cells. The scatter plot shows the macro particles and the solid line the betatron phase advance along the beam.}
	\label{FigFODO}
\end{figure}

We simulate the transport of an electron bunch through a long DLA-based FODO lattice using the aforementioned transfer matrices. Here only the horizontal plane is taken into account. The parameters of the lattice are summarized in Tab.~\ref{tab:lattice} and satisfy the usual stability criterion $f > l_\text{cell} / 4$ for FODO cells of length $l_\text{cell}$ and focal length $f = 1/ks$.
\begin{table}[hbt]
   \centering
   \caption{Parameters of the DLA-based FODO Lattice.}
   \vspace{0.2cm}
   \begin{tabular}{ll}
      \toprule
      \textbf{Parameter} & \textbf{Value}\\
      \midrule
      $N_\text{FODO}$ & 1000\\
      $\beta_\text{m}$ & 0.9999 ($\rightarrow 50\,$MeV)\\
      $\lambda_\text{S}$ & $\beta _\text{m} \cdot $ \SI{2}{\micro\meter}\\
      $N_{\lambda_\text{S}}$ per F,D & 10\\
      $N_{\lambda_\text{S}}$ per O & 100\\
      $G$ & 2.0\,MT/m\\
      \bottomrule
   \end{tabular}
   \label{tab:lattice}
\end{table} 
The result of the transport of a uniform electron beam with 1\,fs length through 1000 FODO cells ($\sim40$\,cm) is shown in Fig.~\ref{FigFODO}. It can be seen that the phase-dependent phase advance indeed has a subtantial effect on the phase space. Note that some parts of the bunch have crossed the channel boundary (dashed line). The effect of this is not taken into account in this study.

	
\section{Conclusion}
We have shown in simulations that it is possible to enhance the equivalent magnetic focusing gradient in simple dual grating type structures by leveraging strong gradients in transient drive laser fields. Furthermore we have shown that it is -- in theory -- possible to adjust both the phase relation between acceleration and focusing and the onset of the enhancement by tailoring the grating geometry. The geometry shown in this paper is not described fully by the simple spatial harmonic model anymore, which predicts a $\pi/2$ phase shift between acceleration and focusing.

In the last section we have shown that a potential simple DLA transport line could suffer from the phase-dependence of the focusing strength $k$ even in the case of short bunches. It has to be investigated if it is possible to mitigate this by looking at more sophisticated transport line configurations. On the other hand it will be interesting to think about potential use cases of the strong phase advance differences along the longitudinal beam slices.

\section{Acknowledgments}
This research is funded by the Gordon and Betty Moore Foundation as part of the Accelerator on a Chip International Program (GBMF4744). 

\section*{References}

\bibliographystyle{elsarticle-num}
\bibliography{\jobname}

\end{document}